\documentclass[11pt]{article}
\newsavebox{\PSLASH}
\sbox{\PSLASH}{$p$\hspace{-1.8mm}/}

\begin{document}
\title{Logarithmic Conformal Null Vectors and SLE }
\author{S. Moghimi-Araghi\footnote{e-mail:
samanimi@sharif.edu}, M. A. Rajabpour \footnote{e-mail:
rajabpour@mehr.sharif.edu}
, S. Rouhani \footnote{e-mail: rouhani@ipm.ir} \\ \\
Department of Physics, Sharif University of Technology,\\ Tehran,
P.O.Box: 11365-9161, Iran} \maketitle

\begin{abstract}

Formal Loewner evolution is connected to  conformal field theory.
In this letter we introduce an extension of Loewner evolution,
which consists of two coupled equations and connect the
martingales of these equations to the null vectors of logarithmic
conformal field theory.
  \vspace{5mm}%
\end{abstract}

\section{Introduction}
In the last two decades, conformal field theory in two dimensions
has found many applications in different areas of physics, such as
string theory, critical phenomena and condensed matter physics. In
particular, the minimal models introduced in \cite{BPZ} reveal
many exact solutions to various two dimensional phase transitions
like Ising model at critical point or three critical Ising model
and so on. In such models, the crucial role is played by the so
called null vectors. In addition these standard transitions,
conformal field theory has shown its strength in another type of
critical phenomena, known as geometric critical phenomena. The
most famous models of this type are perculation and self avoiding
random walks. These are random spatial processes, where either the
probability distribution is determined by equilibrium statistical
mechanics, but main question is about the geometrical properties
of the model (as in perculation), or the probability distribution
is itself geometrical in nature (as in self avoiding walks).

On the other hand, in recent years another way to attack such
problems has been proposed: Stochastic Lowener Evolution (SLE)
\cite{Schramm}. SLE is a probabilistic approach to study scaling
behavior of geometrical models. For a review see
\cite{Rhode,Lawler,kada,kager}. SLE's can be simply stated as
conformally covariant processes, defined on the upper half plane,
which describe the evolution of random domains, called SLE hulls.
These random domains represent critical clusters.

This idea was first developed by Schramm \cite{Schramm}. He showed
that under assumption of conformal invariance, the scaling limit
of self avoiding random walks is SLE$_2$. (SLE's are parameterized
by a real number $k$, and are abbreviated as SLE$_k$) Also he
claimed without proof that SLE$_6$ is the scaling limit of
critical perculation. The claim was proved later on by Smirnov
\cite{Smir}. He showed that in the scaling limit of perculation,
conformal invariance exists and also using the new technic proved
cardy's formula \cite{cardy1,cardy2}.

An explicit relationship between SLE and CFT was discovered by
Bauer and Bernard \cite{bau2,bau3,bau4,bau5}. They coupled SLE$_k$
to boundary conformal field theory with specific central charge
depending on the parameter $k$. These CFT's live on the complement
of SLE hulls in the upper half plane In such situations, boundary
states emerge on the boundary conformal field theory. The good
point is that these states are zero modes of the SLE$_k$
evolution, that is they are conserved in mean. This means that all
components of these states are local martingales of SLE$_k$ and
hence one is able to compute crossing probabilities in purely
algebraic terms. The key point of these result is existence of
null vectors, just an in the case of minimal models.

In this letter we attempt to connect SLE to a certain type of
CFT's, known as logarithmic conformal field theories. These
theories were first introduced by Gurarie \cite{Gur} and have
found many applications in many various areas of physics. For
example see \cite{Flohr,MRSMod} and references therein. The
difference between an LCFT and a CFT, lies in the appearance of
logarithmic as well powers in the singular behavior of the
correlation functions. In an LCFT, degenerate groups of operators
may exist which all have the same conformal weight. They form a
Jordan cell under the action of $L_{0}$. This leads to very
strange properties of these theories.

In this letter we first define SLE$_{k}$ and compactly, then we
review shortly Bauer and Bernard method in connecting SLE$_{k}$ to
CFT. Then we propose a modified version of SLE equation, which in
fact consists of two coupled equations, and connect the two
coupled Loewner's equations to logarithmic conformal field theory.

\section{Chordal Stochastic Loewner Evolutions and CFT}
SLE is a growth process defined via conformal maps which are
solutions of Loewner equation
\begin{equation}\label{SLEMain}
\partial_{t}g_{t}(z)=\frac{2}{g_{t}(z)-\xi_{t}}
\end{equation}
in which $\xi_{t}$ is a  Gaussian random variable obeying the
familiar Langevin's equations of Brownian motion  started from
$B_{t=0}=0$. The autocorrelation of such random variables is
defined to be
\begin{equation}\label{SLEMain1}
E[\xi_{t}\xi_{s}]=k \min(t,s),
\end{equation}
which means that $\xi_t$ is proportional to a normalized Brownian
motion $B_t$, that is : $\xi(t)=\sqrt{k}B_{t}$.

The solution of (1) exists as long as $g_{t}(z)-\xi_{t}$ is non
zero. Let $\tau(z)$ be the first time such that zero is a limit
point of  $g_{t}(z)-\xi_{t}$ as $t\rightarrow\tau$. So, for each
point $z$, the map exists if $t<\tau(z)$. This means that at every
time, only for a subset of the upper half complex plane ($H$) the
map could be defined. We denote these subsets by $K_t$
\begin{equation}
K_t=\left\{z\in H  : \tau(z)\leq t\right\}.
\end{equation}
These subsets are called hulls of SLE evolution. Using these sets
one can define new sets $H_{t}\equiv H/ K_{t}$ (the complement of
$K_{t}$ in upper half plane $H$) which are open are and
conformally equivalent to $H$ via the mapping $g_{t}$. In other
words at each subsequent time t, $g_{t}$ defines a new mapping
region $H_{t}$, which is mapped into $H$. The important property
of $g_{t}$ is that the mapping sets $H_{t}$ continually get
smaller and smaller as t gets larger.  If $\xi(t)$ is continues,
singularities in $g_{t}(z)$ trace out a continues curve in $H$
which are called trace and denoted by $\gamma(t)$. It can be
proved that the trace obeys the equation
\begin{equation}
\gamma(t)= \lim_{\epsilon \rightarrow 0} g_{t}^{-1}(i \epsilon).
\end{equation}
The trace is a path which showes where the singularities arise and
what points have been removed from the mapping region.


To connect SLE$_{k}$ to CFT, it's useful to define new map
$f_{t}(z)\equiv g_{t}(z)-\xi_{t}$ which satisfies the new
stochastic differential equation
\begin{equation}\label{SLEMain2}
df_{t}=\frac{2 dt}{f_{t}}-d\xi_{t}
\end{equation}
The mapping $f_{t}$ belongs to the group $N_{-}$ of germs of
holomorphic functions at $\infty$ and have the form
$z+\sum_{m\leq-1}f_{m}z^{m+1}$. The members of the group of germs
act on themselves: for $f ,g \in N_{-}$ one can define $\gamma_{f}
.g \equiv g \circ f$ with the property $\gamma_{g\circ
f}=\gamma_{f}.\gamma_{g}$. Now we will take an arbitrary function
$F$ in $N_{-}$ and  use Ito's formula  for equation
(\ref{SLEMain2}) to obtain
\begin{equation}\label{SLEMain3}
d\gamma_{f_{t}}.F=(\gamma_{f_{t}}.\acute{F})(\frac{2 dt}{f_{t}}-d
\xi_{t})+\frac{f}{2}(\gamma_{f_{t}}.F^{\prime\prime}).
\end{equation}
By applying $\gamma_{f_{t}}^{-1}$ to both sides of this formula we
have
\begin{equation}\label{SLEMain4}
\gamma_{f_{t}}^{-1}d\gamma_{f_{t}}=2
dt(\frac{2}{z}\partial_{z}+\frac{k}{2}\partial_{z}^{2})-d\xi_{t}\partial_{z}
\end{equation}
The key idea is to associate to any $\gamma_{f}\in N_{-}$ an
operator $G_{f}$ acting on appropriate representation of Virasoro
group. So, the operator $G_{f}$ should satisfy the following
equation
\begin{equation}\label{SLEMain4}
G_{f_{t}}^{-1}dG_{f_{t}}=dt(-2L_{-2}+\frac{k}{2}L_{-1}^{2})+d\xi_{t}L_{-1}.
\end{equation}
Using this formula  Bauer and Bernard have shown how SLE$_{k}$
couples to CFT. They showed if $|\omega\rangle$ be the highest
weight vector of an irreducible Virasoro representation with
central charge $c_{k}=\frac{(3k-8)(6-k)}{2 k}$ and conformal
weight $h_{k}=\frac{6-k}{2 k}$ ,which means it has a singular
vector of level 2 in its descendants, then the time expectation
value $E[G_{f_{t}} |\omega\rangle]$ is a martingale of the
SLE$_{k}$ evolution, that is
\begin{equation}\label{SLEMain5}
E[G_{f_{t}}|\omega\rangle]=G_{f_{s}} |\omega\rangle,
\end{equation}
where the time averaging is for all times less than $s$. This
means that correlation functions of the conformal field theory in
H$_{t}$ are time independent and equal to their value at $t=0$.
Let's what does the state $G_{t}|\omega\rangle$ mean. Suppose
$|\omega\rangle$ be a boundary changing operator in H, then one
can show that the equivalent operator in H$_{t}$ is just
$G_{t}|\omega\rangle$. In fact $G_{t}|\omega\rangle$ is a
generating function for all conserved quantities in chordal SLE.

\section{SLE$_{k}$ and LCFT}

Before establishing a connection between LCFT  and SLE$_k$, we
briefly recall a powerful method introduced in \cite{MRSNil} for
investigating LCFT's. The method is based on a nilpotent weight,
$\theta$, added to the ordinary weight of a primary field. Also a
compound field is defined via
$\Phi(z,\theta)=\phi(z)+\theta\psi(z)$, where $\phi$ and $\psi$
are the usual logarithmic pairs. This new compound field is easily
seen to transform under scaling as
\begin{equation}
\Phi(\lambda z,\theta)=\lambda^{-(h+\theta)}\Phi(z,\theta).
\end{equation}
With such definitions, one is able to extend all the calculations
done in ordinary CFT to LCFT, just by adding a nilpotent variable
where ever the weight of the fields are found.

Following the same scheme, one can generalize Loewner equation in
the following way
\begin{eqnarray}\label{SLELog1}
df_{t}(\theta)=\frac{2dt}{f_{t}(\theta)}-b(\theta)dB_t
\end{eqnarray}
where all the quantities depending on $\theta$ can be expanded in
powers of $\theta$, as an example we have
$f_t(\theta)=f_t^{(1)}+\theta f_t^{(2)}$. The equation itself also
can be expanded in terms of the nilpotent variable so that two
distinct equation is obtained, on for $f^{(1)}$ and the other for
$f^{(2)}$
\begin{eqnarray}\label{SLELog2}
&df^{(1)}_{t}&=\frac{2dt}{f^{(1)}_{t}}-b^{(1)}dB_t\nonumber\\
&df^{(2)}_{t}&=-\frac{2f^{(2)}_{t}dt}{\left(f^{(1)}_{t}\right)^2}-b^{(2)}dB_t
\end{eqnarray}

These two equations are the simplest generalization of the
standard Loewner equation using $\theta$ formalism. Note that the
prefactor of noise should also depend on $\theta$, as it is
related to the weight of the primary field. Also note that the
second equation is coupled to the first one and can not be
considered independently.

Now we should see what happens to the function $F$, which belong
to germs of holomorphic functions. Because we have two distinct
equations, we will have two distinct functions, say $F^{(1)}$ and
$F^{(2)}$, also we have to define two distinct variables $z$ and
$w$ instead of one which happened to be $z$. The variable $z$
evolves with the first equation of (\ref{SLELog2}) and $w$ evolves
with the second equation of (\ref{SLELog2}). In face one can
construct a compound variable $Z(\theta)=z + \theta\, w$ just as
the same we did for $f^{(1)}$ and $f^{(2)}$. It is simply observed
from equations (\ref{SLELog2}) that $F^{(1)}$ just depends on $z$
and $F^{(2)}$ depends both on $z$ and $w$. With a straightforward
calculation, the generalized Ito formula turn out to be
\begin{eqnarray}\label{SLELog3}
dF(\theta)=\left(-2l_{-2}+\frac{1}{2}b(\theta)^{2}l_{-1}^{2}\right)F(\theta)+b(\theta)l_{-1}F(\theta)dB_{t}
\end{eqnarray}
where $l_{n}=-z^{n+1}\frac{\partial}{\partial
z}-(n+1)z^{n}w\frac{\partial}{\partial w}$ satisfy the Witt
algebra $[l_{n},l_{m}]=(m-n)l_{n+m}$. These operators have good
representation in $\theta$ language, just note that
\begin{eqnarray}
\frac{\partial}{\partial z} = \frac{\partial Z(\theta)} {\partial
z}\frac{\partial}{\partial Z(\theta)}= \frac{\partial}{\partial
Z(\theta)},\hspace{1cm} \frac{\partial}{\partial w} =
\frac{\partial Z(\theta)} {\partial w}\frac{\partial}{\partial
Z(\theta)}= \theta\frac{\partial}{\partial Z(\theta)}.
\end{eqnarray}
Now by the definition of $l_n$ we have
$l_n=-\left[Z(\theta)\right]^{n+1}\partial/\partial Z(\theta)$.
 The operators $ l_{n}$ have the counterparts in the Hilbert space of a
conformal field theory, $L_{n}$, which satisfy the well known
Virasoro algebra. So, just as in the case of ordinary CFT, we can
define proper Virasoro transformations associated with these new
functions. Theses Virasoro operators should satisfy the following
equation
\begin{eqnarray}\label{SLELog4}
dG(\theta)=G(\theta)\left(-2L_{-2}+\frac{1}{2}b(\theta)^{2}L_{-1}^{2}\right)
+G(\theta)b(\theta)L_{-1}dB_{t},
\end{eqnarray}
which is again just like the one derived in CFT case, of course
with $\theta$ modifications.

before stating the connection between LCFT and the generalized SLE
equation, we briefly recall how singular vectors are investigated
in the context of LCFT's. The problem of singular vectors in
LCFT's have been considered in several papers
\cite{Fl-Sing,MRSNil}, we will follow mainly \cite{MRSNil} as the
language is close to the one we have used here. In an LCFT, the
representation of Virasoro algebra is constructed from a compound
highest weight vector $|\omega+\theta\rangle= |\omega^{(1)}\rangle
+ \theta\, |\omega^{(2)}\rangle$, with the properties
\begin{eqnarray}
&&L_0|\omega+\theta\rangle=(h+\theta)|\omega+\theta\rangle,\nonumber\\
&&L_n|\omega+\theta\rangle=0\hspace{2.4cm}(n\geq 1).
\end{eqnarray}
This means that $|\omega^{(1)}\rangle$ and $ |\omega^{(2)}\rangle$
have the same weight and form a $2\times 2$ Jordan cell. All the
descendants are produced by applying $L_{-n}$'s to these states or
equivalently to $|\omega+\theta\rangle$ with $n>0$. There may be
some representations, in which some of the descendants are
perpendicular to all other vectors including themselves. These are
the singular vectors. As an example with proper adjustment of $h$,
the weight of the primary field and the central charge, the vector
\begin{equation}\label{Nul}
|\chi(\theta)\rangle=\left(-2L_{-2}+\frac{3}{2(h+\theta)+1}L_{-1}^{2}\right)
|\omega+\theta\rangle
\end{equation}
turns out to be a singular one. If you compare the term appearing
in parenthesis with the one obtained in equation (\ref{SLELog4}),
you'll find a fascinating similarity, just take
$b(\theta)^{2}=\frac{3}{2(h+\theta)+1}$ then, you will have
completely the same combination (We have dropped the last term in
(\ref{SLELog4}) for the moment). Let's operate on both sides of
(\ref{SLELog4}) with $|\omega+\theta\rangle$.  gathering all the
terms we arrive at
\begin{equation}\label{SLELog5}
dG(\theta)|\omega+\theta\rangle=G(\theta)b(\theta)L_{-1}dB_{t}|\omega+\theta\rangle.
\end{equation}
From the definition of Ito integrals, $G(\theta)$ is independent
of $dB_t$, so the time average of the right hand side is vanishes.
This means that the time average of changes in
$G(\theta)|\omega+\theta\rangle$ is zero and it is a martingale of
the model.

{\large \bf Acknowledgement}: We would like to thank M. Saadat and
M. R. Rahimi-Tabar for their helpful discussions.


\begin{thebibliography}{99}
\bibitem{BPZ}   A. Belavin, A. Polyakov and A. Zamolodchikov, Nucl. Phys. \textbf{B241}(1984)333
\bibitem {Schramm}  O.Schramm, Israel, J. Math. 118 (2000) 221.
\bibitem {Rhode}     S. Rhode, O. Schramm, [Math.PR/0106036]
\bibitem {Lawler}     Lawler, An introductory text, the draft of book may
be found at http://www.math.cornel.edu/~Lawler
\bibitem {kada}    I. A. Gruzberg and L. kadanoff,
[cond-mat/0309292] and refrences therein.
\bibitem{kager} W. Kager and B. Nienhuise, [math-Ph/0312030]
\bibitem{Smir} S. Smirnov, C. R. Acad. Sci. Paris Sr. I Math
{\bf 333}, 239, 2001
\bibitem {cardy1}     J. Cardy, [math-Ph/0103018]
\bibitem {cardy2}      J. Cardy, J. Phys. A 25(1992)L201.
\bibitem {bau2} M. Bauer and D. Bernard, Phys. Lett. B 543(2002)135-138
[math-Ph/0206028]
\bibitem {bau3}  M. Bauer and D. Bernard, Comm. Math. Phys. 239: 493-521,
2003. [hep-th/0210015]
\bibitem {bau4}     M. Bauer and D. Bernard, Phys. Lett. B 557: 309-316, 2003.
[hep-th/0301064]
\bibitem {bau5}     M. Bauer and D. Bernard, [math-Ph/0305061]
\bibitem{Gur}  V.Gurarie, Nucl. Phys. \textbf{B410} (1993) 535 [hep-th/9303160]
\bibitem{Flohr}  M. Flohr, Int. J. Mod. Phys. A18 (2003)
4497-4592.[hep-th/0111228]
\bibitem{MRSMod} Int. J. Mod. Phys. A18
(2003) 4747-4770 [hep-th/0201099]
\bibitem {Fl-Sing} M. Flohr, Nucl. Phys. {\bf B514} (1998) 523
[hep-th/9707090]
\bibitem{MRSNil} S. Moghimi-Araghi, S. Rouhani and M. Saadat, Nucl.Phys.
\textbf{B599} (2001) 531-546, [hep-th/0008165]
\end{thebibliography}
\end{document}